\documentclass[aps,prl,floatfix,twocolumn,groupedaddress]{revtex4}

\usepackage{graphicx}

\begin{document}

\title{Electromagnetic turbulence driven by the mixed mode instability}

\author{Jacob Trier Frederiksen}
\affiliation{Niels Bohr Institute, Juliane Maries Vej 30, 2100 K\o benhavn \O, Denmark}
\email[]{trier@astro.ku.dk}

\author{Mark Eric Dieckmann}
\affiliation{Centre for Plasma Physics, Queen's University Belfast, U.K.}
\altaffiliation{Department of Science and Technology (ITN), Link\"oping 
University, SE-601 74 Norrk\"oping, Sweden}

\date{\today}

\begin{abstract}
In continuation of previous work, numerical results are presented, concerning relativistically counter-streaming plasmas. Here, the relativistic mixed mode instability evolves through, and beyond, the linear saturation -- well into the nonlinear regime. Besides confirming earlier findings, that wave power initially peaks on the mixed mode branch, it is observed that, during late time evolution wave power is transferred to other wave numbers. It is argued that the isotropization of power in wavenumber space may be a consequence of weak turbulence. Further, some modifications to the ideal weak turbulence limit is observed. Development of almost isotropic predominantly electrostatic -- partially electromagnetic -- turbulent spectra holds relevance when considering the spectral emission signatures of the plasma, namely bremsstrahlung, respectively magneto-bremsstrahlung (synchrotron radiation and jitter radiation) from relativistic shocks in astrophysical jets and shocks from gamma-ray bursts and active galactic nuclei.
\end{abstract}

\maketitle

Counter-streaming plasmas subject to the relativistic mixed mode instability (MMI), driven by relativistic beams of electrons and ions, have previously been examined theoretically \cite{Bret,Bret2}, and numerically with particle-in-cell (PIC) simulations \cite{Dieckmann,Gremillet}. Growth rates of the mixed mode, the filamentation and the two-stream instabilities (hereafter MMI, FI and TSI, resp.), are 
\begin{eqnarray*}
	\gamma_{MM} & = & \sqrt{3}(n_b/16n_p)^{1/3}\Gamma(v_b)^{-1/3} \\
	\gamma_{F}  & = &  (v_b/c)(n_b/  n_p)^{1/2}\Gamma(v_b)^{-1/2} \\
	\gamma_{TS} & = & \sqrt{3}(n_b/16n_p)^{1/3}\Gamma(v_b)^{-1}~~,
\end{eqnarray*}
with $n_b$, $n_p$, $v_b$ and $\Gamma (v_b) \equiv {(1-v_b^2/c^2)}^{-1/2}$, the beam and background plasma density, the beam speed and beam bulk Lorentz factor, respectively, in the background rest frame. For a large volume in parameter space, $\mathcal{V}\in\{n_b,n_p,v_b,\Gamma(v_b)\}$, the MMI will have the highest growth rate of the three possible instabilities for the relativistic beam-plasma interaction. The MMI propagates at oblique angles with respect to the beam velocity $\angle(\mathbf{k}_{MMI},\mathbf{v}_b) \propto \arctan{[(v_b/v_p)^{1/2}]}$, where $v_p$ is the background plasma thermal speed and $v_p \ll v_b$ \cite{Bret}. A thorough theoretical discussion of the relationship between the MMI, FI and TSI modes is given in \cite{Bret2}.

Due to its mixed nature, the MMI contains both an electrostatic and an electromagnetic wave component \cite{Bret2}. Potentially, both electrostatic and electromagnetic turbulence (wave mode coupling leading to cascades/inverse cascades in $k$-space) is possible in such systems. This potential for producing very broad band plasma turbulence (in both $\textbf{E}$ and $\textbf{B}$ fields) is highly relevant when considering for example inertial confinement fusion experiments \cite{Badziak,Tabak}. Other examples where electromagntic wave turbulence is important are astrophysical jets and shocks from gamma-ray bursts and active galactic nuclei, where ambient plasma streams through a shock interface moving at relativistic speeds -- see e.g.~\cite{Frederiksen,Medvedev,Hededal}. Further studying of the MMI in particular, here, beyond the linear regime to the development of turbulence is thus highly motivated.

Performing high resolution 2.5D PIC code simulations, we examine the turbulent wave spectra, that develop during saturation of the MMI by the trapping of electrons \cite{Dieckmann,Gremillet}. The code solves Lorentz' equations of motion for an ensemble of computational macro-particles (CPs) defined in 2D3V; $\{x,y,p_x,p_y,p_z,t\}$, and Maxwell's equations in two spatial dimensions; $\{\textbf{E}_{x,y},\textbf{B}_{z}\}(x,y,t)$.  A PIC code scaling is chosen wherein $m_e = e = c \equiv 1$ for the electron mass, elementary charge and light speed. 

We model a relativistic beam consisting of co-moving electrons and protons with a mass ratio $m_i/m_e = 1836$ and equal number densities. The background plasma consists also of equal number densities of electrons and protons. The computational system is consequently charge neutral and current neutral and neither electric, nor magnetic, fields are present, initially. The background plasma is chosen as our laboratory frame. Here, the bulk flow velocity vector of beam electrons and beam protons is set to $\mathbf{v}_b = v_b \hat{\mathbf{y}} \approx 0.97c~\hat{\mathbf{y}}$, i.e. with $\Gamma (v_b) \equiv 4$. Densities and temperatures, for the cold beam and warm background plasmas are set to $n_b = 0.1$ and $T_b = 10^{-4}$ ($v_{th,b}=0.01c$), and $n_p = 0.9$ and $T_p = 10^{-2}$ ($v_{th,p}=0.1c$), using the aforementioned scaling of the relevant natural constants. With these parameter choices the plasma frequency is $\omega_{pe} \approx 0.9$, and the electron skin depth becomes $\delta_e = c / \omega_{pe} \approx 1.1$. Time is henceforth normalized to the plasma frequency, $t = t' \omega_{pe}$, and wavenumbers to the volume dimensions, i.e. $k_x=1=2\pi/L_x$. All boundary conditions on both the fields and particles are periodic. 

\begin{figure*}
\begin{center}
\includegraphics[width=\textwidth]{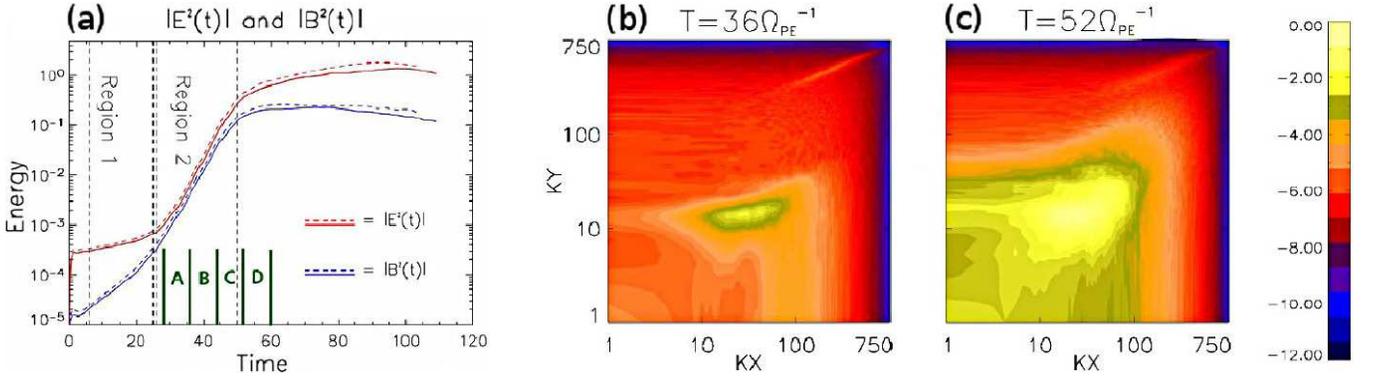}
\caption{(Color) (a) The electric and magnetic energy 
densities $e_E$ (red upper curves) and $e_B$ (blue lower curves) measured 
for Simulation 1 (dashed curves) and Simulation 2 (solid curves). 
Noise builds up in the region 1 and the MMI develops linearly in 
the region 2. The ratio $e_E/e_B$ increases after the saturation, at $t \omega_{e,1} \approx 50$. We evaluate the time evolution of wave power spectra 
on intervals A through D. (b) the power spectrum, $P(k_x,k_y,t=36)$, corresponds to spectrum of the MMI during linear growth. (c) shows the same spectrum just after saturation, $P(k_x,k_y,t=52)$. $k_{x,y}$ are in units of $2\pi / L_{x,y}$. The color scale is 10-logarithmic and identical for both (b) and (c).}
\label{Fig1.1}
\end{center}
\end{figure*}

To ensure robustness and numerical accuracy of our results we have compared two simulations with the same physical parameters but with different grid and particle resolutions. \textsl{Simulation 1} uses a grid size $L_x=1500\Delta_x=74\delta_e$, $L_y=1500\Delta_y=74\delta_e$ where $\Delta_x=\Delta_y$ is the grid spacing. The plasma is resolved in particle phase space with $N_e=N_p=200$~CPs/cell/species -- a total of $N_t = 400$~CPs/cell. In \textsl{Simulation 2} the corresponding run parameters are: $L_x=59\delta_e=1200\Delta_x$, $L_y=59\delta_e=1200\Delta_y$, $\Delta_x = \Delta_y$ and $N_t = 600$~CPs/cell. Figure~\ref{Fig1.1}(a) displays the electric and magnetic energy densities, integrated over the simulation volume. The data are normalized to the total electric field energy for Simulation 1 at $t=t_{end}=110\omega_{pe}^{-1}$. As seen from figure~\ref{Fig1.1}(a) we obtain excellent agreement for the development of the energy densities through all stages of the MMI instability in the reference run. The small discrepancy in energy is solely due to a slightly higher initial noise level of Simulation 2 that persists throughout; this, in turn, shows us that energy conservation is also agreeable.  Having checked that all other results presented here are also robustly reproduced in Simulation 2, we present data solely from Simulation 1. We avoid re-examination of the electron distributions and refer to earlier work treating the linear MMI phase~\cite{Dieckmann,Gremillet} and dynamics of the ions. We merely note here that the ions begin to undergo instability in the late linear MMI phase and has important influence on long time evolution. The remainder of this brief communication focuses on the field dynamics during transition to non-linear development. 

Figure~\ref{Fig1.1}(a) displays the total field energy history. Initially, fields are due to noise and practically electrostatic already after $t\sim5$ (Region 1 in Fig.\ref{Fig1.1}(a)~). Magnetic fluctuations, however, rapidly build up to equipatition with the electric fluctuations. Once in equilibrium, the ratio $E_B$ grows in unison with $E_E$ through the late linear stage (Region 2) of the MMI, albeit at a slightly lower value. This ratio again increases after MMI saturation at $t>50$. It reaches a quasi-steady value $E_E/E_B\approx 10$ at the simulation's end ($60 < t < t_{end}\equiv110$), signifying the predominance of electrostatic fields.

\begin{figure*}
\begin{center}
\includegraphics[width=\textwidth]{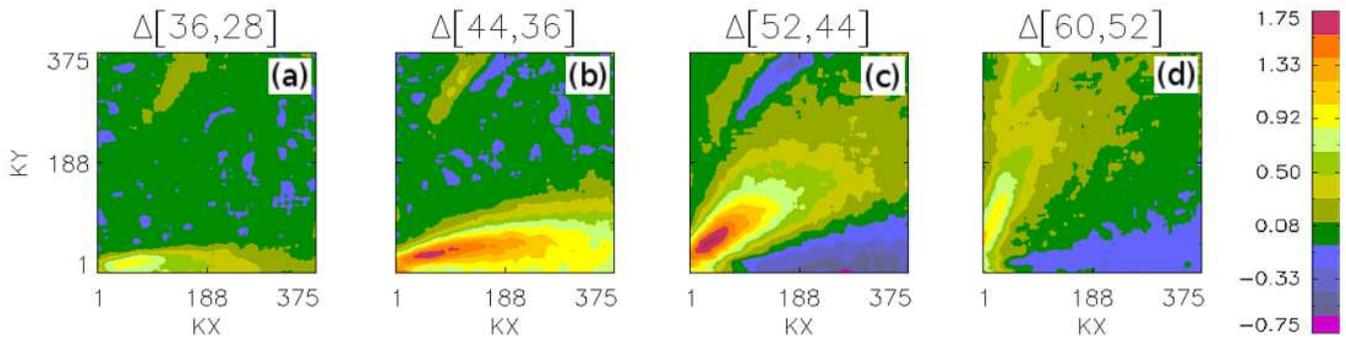}
\caption{(Color) Temporal development of the power spectrum ratio, $R(k_x,k_y,t,\Delta_t)$, at the
times $t_1 \omega_{e,1}=$ 36, 44, 52, 60 -- (a), (b), (c) and (d) respectively.
Color scaling is 10-logarithmic, and zero-level (blue) equals zero growth rate; i.e. Log$_{10}(R(k_x,k_y,t,\Delta_t)) \equiv 0$, following from a ratio $P(t_1)/P(t_2) \equiv 1$. Only one quarter, $\{1<k_x<375,1<k_y<375\}$, of the full spectrum quadrant, for $\{k_x>0,k_y>0\}$, is shown for purposes of visual clarity. Note the $k$-scale is linear.}\label{Fig1.2}
\end{center}
\end{figure*}

Figure~\ref{Fig1.1}(b,c) displays power spectra $P_E(k_x,k_y,t)=|FFT[E_x(t) + iE_y(t)]|^2$ of the electric field, in the quadrant $\{k_x,k_y\} > \{0,0\}$ at times $t=36$ (during linear MMI) and $t=52$ (just prior to saturation). 
We have 
rendered only a single quadrant for the sake of clarity and display resolution. The spectral representation of the electric field, $P_E(k_x,k_y,t)$ reveals the unstable wave branch, and the $\mathbf{k}$-range of the turbulent spectrum.  With $\delta_{e} \simeq 1.1$ and $v_b \approx 0.97 c$ (from $\Gamma(v_b) = 4$), we find the most unstable mode at $\tilde{k}_y \approx (0.98\delta_e)^{-1}$ or $k_y = \tilde{k}_y / (2\pi / 74 \delta_e) \approx 11$, which agrees well the $k_y$ of the MMI branch in Fig.\ref{Fig1.1}(b,c). Significant wave power is present over a broadened range of $k_x,k_y$ at $t\omega_{e,1}=52$, which peaks on the initial MMI branch at $t\lesssim36$. The magnetic power spectrum $P_B(k_x,k_y,t)=|FFT[B_z(t)]|^2$ (not shown) -- also due to the MMI \cite{Bret2} -- follows closely that of the electric field albeit at lower power until about $t\sim50$ when linear phase has saturated. Also, power is slightly enhanced at large $k_x,k_y$ due to the finite grid instability, which was treated in \cite{Dieckmann}. 

Figure~\ref{Fig1.1}(b,c) also reveals a rapid time dependent spreading of wave power away from the mixed mode branch. This effect is not due to Larmor effects of the magnetic field; even if assuming a peak magnetic field $B_{max} \approx 0.2$ (measured) everywhere during $20 < t < 60$, yields a relativistic electron gyro frequency of $\omega_{ce} \approx \omega_{pe}(eB_{max})/(m_e\Gamma(v_b)c) \approx 0.05\omega_{pe}$. Hence, the Larmor time scale is not a governing time scale.
Langmuir wave turbulence has been found~\cite{Ziebell} to spread into a ring distribution power spectrum in $\mathbf{k}$-space, assuming a weak turbulence model. This result was obtained for initially mono-directional Langmuir electrostatic turbulence, in the non-relativistic limit~\cite{Ziebell}. 
Our simulations -- on the other hand -- fully resolve all particle and wave dynamics. Consequently, we capture the spectral dynamics of both weak turbulence (wave-wave interaction) and strong turbulence (wave-particle interaction).
By analysing the rapid temporal evolution of $P_E(k_x,k_y,t)$, we can determine whether the observed spread of the power in $\mathbf{k}$-space is partially or fully due to weak Langmuir turbulence.  The $\mathbf{k}$-interval with the strongest change should rotate around $k_x,k_y = 0$ on shells of $|k|=const$.  The quantity, of interest is $$R(k_x,k_y,t_1,\Delta_t) = \log_{10} (P_E[k_x,k_y,t_1]/P_E[k_x,k_y,t_0])~,$$ which captures the \textit{difference} of $P_E(k_x,k_y,t)$, at two successive simulation time steps. We calculate $R(k_x,k_y,t_1,\Delta_t)$ at four times $t_1=36,44,52$ and $60$, with $\Delta_t = (t_1-t_0)=8$, corresponding to the time intervals "A" through "D" in Fig.\ref{Fig1.1}(a).

Indeed, figure~\ref{Fig1.2} reveals rich structure and dynamics of $R(k_x,k_y,t_1,\Delta_t)$. Initially ($t<36$) power grows uniformly, covering a range, $1<k_x<280$, and $1<k_y<50$, mainly along $k_x$-axis. Later it rotates away from the $k_x$-axis, towards the axis $k_y$-axis while maintaining its elongated shape. Power is transferred between the $x$- and $y$-directions on curves of approximately constant $|k|$. We conclude that, qualitatively, what is observed from Fig.\ref{Fig1.2}(a-d) is a consequence of driven weak turbulence. Some modification to pure ($k_x$,$k_y$)-rotation results from relaxing the  assumptions of~\cite{Ziebell}. A much weaker structure disperses (both positive (light green) and negative (blue) growth) wave power produced on the finite grid instability~\cite{Dieckmann} branch. It is eventually outpowered by MMI driven turbulence.

Ideal weak turbulence scatters wave energy~\cite{Ziebell} on shells azimuthally centered in $k$-space ($k_x$,$k_y$). 
On the other hand strong turbulence, mediated by wave-particle interactions and phase space hole coalescence, transfers wave energy to lower wavenumbers along $k_y$, hence perpendicular to the wave scattering direction of weak turbulence (which occurs on constant $k$-shells). Wave-particle interactions in mono-directional turbulence were considered previously~\cite{Schlickeiser,Fleishman1}, as was the interplay between electrons and isotropic Langmuir turbulence~\cite{Fleishman2}. The MMI in our results, however, furthermore yields turbulent magnetic fields, which were not taken into account by~\cite{Schlickeiser,Fleishman1,Fleishman2}.  At late times the magnetic field will participate increasingly in producing particle trapping and contribute magneto-bremsstrahlung (synchrotron/jitter~\cite{Medvedev}) to the radiation signature of the plasma.

\begin{figure*}
\begin{center}
\includegraphics[width=\textwidth]{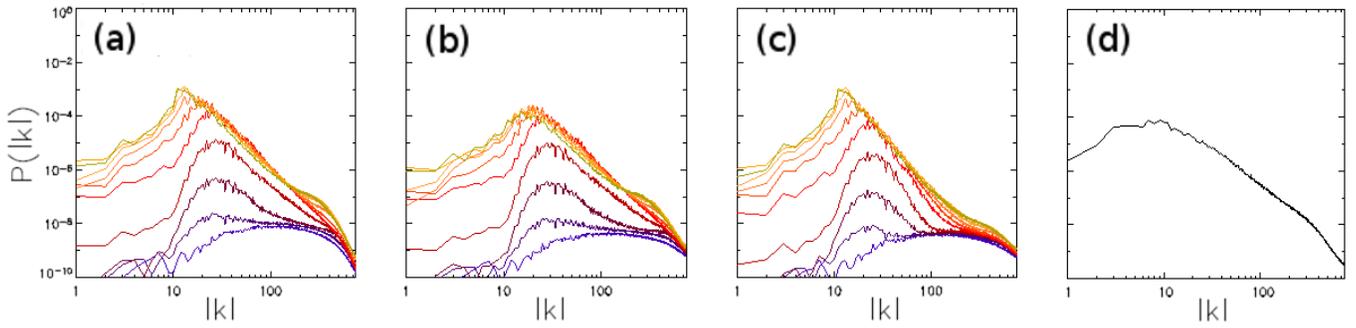}
\caption{(Color online) Power spectra for the total, $P(\left|E_x(k) + i~E_y(k)\right|^2)$, the transverse, $P(E_x(k))$, and the longitudinal, $P(E_y(k))$, electric field components -- (a), (b) and (c) respectively. The spectra are integrated azimuthally, $\phi\in[0,\pi/2]$, over shells in $k$-space, $\{k_x,k_y\}=\{|k|\cos\phi,|k|\sin\phi\}$ -- cf. fig.\ref{Fig1.1}(b,c). Each curve correponds to one time instance, and these are distributed at equal intervals, $t=[0,10..,100,110]$. Curves of low power (blue) gives the spectra for early times, and curves of high power (orange and yellow) reveils late time spectra. (d) Power spectrum of the total (out-of-plane) magnetic field, $P(B_z(k))$, shown only for simulation end time, $t = t_{end} = 110$. The total magnetic-to-electric energy content at $t_{end}$ is $E_B/E_E \approx 10$; a considerable amount of electromagnetic power is present but is still less than the electrostatic power.}
\label{Fig1.3}
\end{center}
\end{figure*}

Lastly, figure~\ref{Fig1.3}(a-c) displays azimuthally integrated power spectra of the electric field for 10 different timesteps during the simulation. Furthermore, for a single time, $t=t_{end}\equiv110$, and Fig.\ref{Fig1.3}(d) shows the corresponding spectrum for the magnetic field. At the earliest times (low power) we observe, again, growth of small scale noise which is seen to be approximately thermal. This noise is generic to PIC codes. However, Fig.\ref{Fig1.3}(b), $P(E_x(k_\phi))$), we see that the high $k$ noise grows further at all times. This we attribute mainly to the finite grid instability in the early stages. At later times channeling of power from low $|k|$ to high $|k|$ is due to the steepening of structures by nonlinear effects \cite{Rowlands}; turbulence reaches $k$-scales comparable to the inverse Debye length. Both these effects at high $|k|$ are electrostatic which is seen by comparing with Fig.\ref{Fig1.3}(d); the magnetic field contains negligable power in deviation from that of a power-law on the inertial range. \\

In summary, we have examined -- numerically -- turbulent wave spectra driven by the mixed mode instability (MMI) in 2D3V PIC simulations. Parameters for the simulations were chosen so as to favor the MMI over both the FI and the TSI.  

We interpret our main result as a consequence of weak turbulence, which transfers power turbulently on shells of constant wavenumbers~\cite{Ziebell} as seen in Fig.\ref{Fig1.2}(a-d). Neither the coalescence of filaments -- transferring power to lower $k_x$~\cite{Morse,Rowlands}, nor phase space hole coalescence -- transferring power to lower $k_y$~\cite{Berk,Schamel,Bengt}, however, can be captured in an ideal weak turbulence model. This may explain why we do not observe a neither simple ring distribution, nor constancy of the structure in Fig.\ref{Fig1.1}(a), along $\phi = const.$

Future work should be conducted to determine the relative strength of the (ideally) weak and strong turbulence contributions to the total dynamics. In particular aspects of strong turbulence deserve thorough investigation. Detailed analysis of the powerlaw slopes in Fig.\ref{Fig1.3}(a-d) is also important, for determination of the relative strengths of magneto-bremsstrahlung (synchrotron/jitter~\cite{Medvedev}) and bremsstrahlung spectral contributions.

A thorough parametric scan and a characterization of electromagnetic turbulence in shocks is important, when linking radiation spectra with predictions of physical conditions in a broad class of astrophysical jets. \\

\begin{acknowledgments}

We acknowledge financial support from Danish Natural Science Research Council (JTF), 
the Swedish Research Council (MED,JTF), and by a UK EPSRC Science and Innovation Award (MED). The Danish Center for Scientific Computing (DCSC) provided computer time and support. 

\end{acknowledgments}


~ \\ ~


\end{document}